\documentclass[aps,prb,twocolumn,superscriptaddress]{revtex4-1}

\usepackage{epsfig}
\usepackage{amssymb} 
\usepackage{amsfonts} 
\usepackage{mathtools} 
\usepackage{dcolumn} 
\usepackage{graphicx} 
\usepackage{color}  
\usepackage{bm}  
\usepackage{comment} 
\usepackage{units}
\usepackage{sidecap} 
\usepackage{textcomp} 
\usepackage{units}
\usepackage{mathrsfs} 
\usepackage{epsfig}
\usepackage{amssymb} 
\usepackage{amsfonts} 
\usepackage{mathtools} 
\usepackage{dcolumn} 
\usepackage{graphicx} 
\usepackage{color}  
\usepackage{bm}  
\usepackage{units}
\usepackage{sidecap} 
\usepackage{textcomp} 
\usepackage{mathrsfs} 

\usepackage{wrapfig} 
\usepackage{array} 
\usepackage{upgreek}
\usepackage{tabularx}
\usepackage{latexsym}
\usepackage{bm,array}
\usepackage{amsfonts}
\usepackage{amssymb}
\usepackage{amsmath}
\usepackage{bigstrut}
\usepackage{placeins}

\newcommand{\bpm}{\begin{pmatrix}}
\newcommand{\epm}{\end{pmatrix}}
\newcommand{\be}{\begin{eqnarray}}
\newcommand{\ee}{\end{eqnarray}}
\newcommand{\ba}{\begin{array}}
\newcommand{\ea}{\end{array}}
\def\F{
\begin{bmatrix}
    1 \\
    1 \\
    1 \\
    1
\end{bmatrix}}
\def\C{
\begin{bmatrix}
     \,\,\,\,1 \\
     \,\,\,\,1 \\
    -1 \\
    -1
\end{bmatrix}}
\def\A{
\begin{bmatrix}
    \,\,\,\,1 \\
    -1 \\
    -1 \\
    \,\,\,\,1
\end{bmatrix}}
\def\G{
\begin{bmatrix}
    \,\,\,\,1 \\
    -1 \\
    \,\,\,\,1 \\
    -1
\end{bmatrix}}

\usepackage{multirow}

\def \anio{$\alpha-$Na$_2$IrO$_3$}
\def \alio{$\alpha-$Li$_2$IrO$_3$}
\def \blio{ $\beta-$Li$_2$IrO$_3$}
\def \glio{ $\gamma-$Li$_2$IrO$_3$}
\usepackage{array} 
\usepackage{upgreek}

\def \anio{$\alpha$-Na$_2$IrO$_3$}
\def \alio{$\alpha$-Li$_2$IrO$_3$}
\def \blio{$\beta$-Li$_2$IrO$_3$}
\def \glio{$\gamma$-Li$_2$IrO$_3$}
\def\F{
\begin{bmatrix}
    1 \\
    1 \\
    1 \\
    1
\end{bmatrix}}
\def\C{
\begin{bmatrix}
     \,\,\,\,1 \\
     \,\,\,\,1 \\
    -1 \\
    -1
\end{bmatrix}}
\def\A{
\begin{bmatrix}
    \,\,\,\,1 \\
    -1 \\
    -1 \\
    \,\,\,\,1
\end{bmatrix}}
\def\G{
\begin{bmatrix}
    \,\,\,\,1 \\
    -1 \\
    \,\,\,\,1 \\
    -1
\end{bmatrix}}

\begin{document}
\title{Field-induced intertwined orders in 3D Mott-Kitaev honeycomb \blio}

\author{Alejandro Ruiz}
\affiliation{Department of Physics, University of California, Berkeley, California 94720, USA}
\affiliation{Materials Sciences Division, Lawrence Berkeley National Laboratory, Berkeley, California 94720, USA}

\author{Alex Frano}
\affiliation{Department of Physics, University of California, Berkeley, California 94720, USA}
\affiliation{Advanced Light Source, Lawrence Berkeley National Laboratory, Berkeley, California 94720, USA}

\author{Nicholas P. Breznay}
\affiliation{Department of Physics, University of California, Berkeley, California 94720, USA}
\affiliation{Materials Sciences Division, Lawrence Berkeley National Laboratory, Berkeley, California 94720, USA}

\author{Itamar Kimchi}
\affiliation{Department of Physics, University of California, Berkeley, California 94720, USA}
\affiliation{Department of Physics, Massachusetts Institute of Technology, Cambridge, MA 02139, USA}

\author{Toni Helm}
\affiliation{Department of Physics, University of California, Berkeley, California 94720, USA}
\affiliation{Materials Sciences Division, Lawrence Berkeley National Laboratory, Berkeley, California 94720, USA}

\author{Iain Oswald}
\affiliation{Department of Chemistry, The University of Texas at Dallas, Richardson, Texas 75080, USA}
\author{Julia Y. Chan}
\affiliation{Department of Chemistry, The University of Texas at Dallas, Richardson, Texas 75080, USA}

\author{R. J. Birgeneau}
\affiliation{Department of Physics, University of California, Berkeley, California 94720, USA}
\affiliation{Materials Sciences Division, Lawrence Berkeley National Laboratory, Berkeley, California 94720, USA}

\author{Z. Islam}
\affiliation{Advanced Photon Source, Argonne National Laboratory, Argonne, Illinois 60439, USA}

\author{James G. Analytis}
\affiliation{Department of Physics, University of California, Berkeley, California 94720, USA}
\affiliation{Materials Sciences Division, Lawrence Berkeley National Laboratory, Berkeley, California 94720, USA}

\date{\today}

\begin{abstract}

Honeycomb iridates are thought to have strongly spin-anisotropic exchange interactions that could lead to an extraordinary state of matter known as the Kitaev quantum spin liquid. The realization of this state requires almost perfectly frustrated interactions between the magnetic Ir$^{4+}$ ions, but small imbalances in energy make other ordered states more favorable. Indeed, the closeness in energy of these ordered states is itself a signature of the intrinsic frustration in the system. In this work, we illustrate that small magnetic fields can be employed to drive the frustrated quantum magnet \blio\,between different broken symmetry states, but without causing a true thermodynamic phase transition. This field-induced broken symmetry phase has all the signatures of a thermodynamic order parameter, but it is never truly formed in zero field. Rather, it is summoned when the scales of frustration are appropriately tipped, intertwined with other nearby quantum states.

\end{abstract}


\maketitle

Materials with nearly degenerate ground states are arguably at the center of some of the most complex and technologically important problems in condensed matter physics. Degeneracies are the reason that strongly correlated materials have rich phase diagrams~\cite{witczak-krempa_interacting_2014}, can be the origin of topological defects~\cite{berry_quantal_1984}, and are thought to be crucial for the realization of a fault-tolerant quantum computer~\cite{kitaev_anyons_2006}. One class of such materials are known as quantum spin liquids (QSL). In theory, they are hosts for novel states of matter, generally arising from strong magnetic frustration and characterized by highly degenerate excitations~\cite{kitaev_anyons_2006, balents_spin_2010}. Mott-insulator iridates have ignited interest in this field partially due to their theoretical connection to an exactly soluble QSL discussed by Kitaev~\cite{kitaev_anyons_2006,jackeli_mott_2009}, a critical ingredient of which is that every magnetic ion in a honeycomb lattice couples an orthogonal component of spin to each of its three nearest neighbors. 

The ``Mott-Kitaev" iridates crystallize in 2D and 3D structures (the harmonic honeycombs~\cite{modic_lio}), and all known compounds are found to magnetically order at low temperature in one of two ways: a commensurate `zig-zag' phase (found in \anio\,and $\alpha$-RuCl$_3$~\cite{ye_direct_2012, choi_spin_2012,sears_magnetic_2015}), and an incommensurate order (found in $\alpha$, $\beta$ and \glio~\cite{modic_lio, takayama_hyperhoneycomb_2015, williams_incommensurate_2016}). This suggests that there are non-Kitaev interactions in the Hamiltonian, relieving the frustration and obscuring any low-energy signature of the Kitaev physics~\cite{kimchi_unified_2015, jackeli_mott_2009, lee_theory_2015}. As a result, the presence of Kitaev-like terms is often inferred from scattering studies~\cite{banerjee_proximate_2016, hwan_chun_direct_2015} or from evidence of anomalous dissipative processes in spectroscopic measurements~\cite{knolle_raman_2014, gupta_raman_2016}. A central question in these materials is whether these different ordered states are manifestations of the same physics, and in particular whether this physics is dominated by Kitaev-like interactions.

\begin{figure*}[ht]
    	{\includegraphics[width=2\columnwidth]{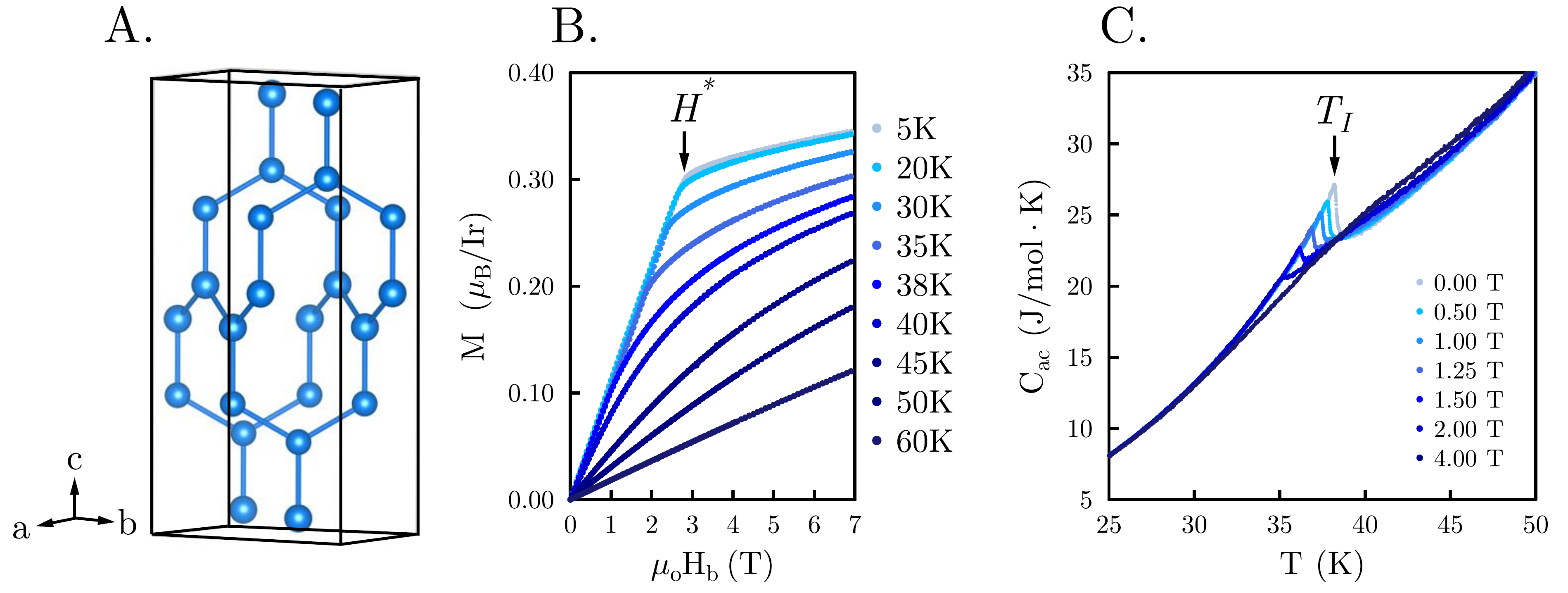}
  	\caption{ {\bf Thermodynamic properties of single crystal $\beta$-Li$_2$IrO$_3$ with magnetic field along the $\hat{b}$-axis.} (A) 3D projection of a unit cell. The Ir atoms (blue balls) form zig-zag chains stacked along the $\hat{c}$-axis and directed alternatingly along the $\hat{a}\pm\hat{b}$ directions. (B) Magnetization versus magnetic field for temperatures up to $\unit[60]{K}$. The low temperature data shows a kink at $H^*$. (C) Heat capacity as a function of temperature taken at magnetic fields up to $\unit[4]{T}$.} 
  	\label{fig:susc}}
	\end{figure*}

In this work, we reveal evidence for nearly degenerate broken symmetry states in \blio, a signature of the underlying magnetic frustration. This compound is the simplest of the 3D harmonic honeycomb structures, composed of interwoven networks of hexagonal chains propagating in the $\hat{a}\pm\hat{b}$ directions (Fig.~\ref{fig:susc}A). Importantly, Kitaev exchange along the $\hat{c}$-axis bonds should couple spins pointing in the $\hat{b}$-axis, making the $\hat{b}$-axis magnetically special, and thus we focus on the response of the system to an applied field in this direction. We find that in this configuration, the `zig-zag' order familiar from other honeycomb compounds can be summoned into existence at the expense of the incommensurate order, identifying an intriguing familial connection between different Mott-Kitaev structures.

{\bf Experimental Results.} Single crystals of $\beta$-Li$_2$IrO$_3$ were synthesized using a vapor transport technique as described in Methods. Fig.~\ref{fig:susc}B,C shows the thermodynamic response of a single crystal with a magnetic field applied along the $\hat{b}$-axis. The temperature-dependent heat capacity at various fixed magnetic fields is shown in Fig.~\ref{fig:susc}C. The anomaly at $T_I=\unit[38]{K}$, ($\mu_0H=\unit[0]{T}$)  corresponds to the onset of a known incommensurate magnetic structure with order parameter $\Psi_{\textrm{I}}$ and propagation vector $q=(0.574,0,0)$~\cite{biffin_noncoplanar_2014}. The anomaly is highly triangular, indicative of a mean-field, second order transition. This feature is strongly suppressed by magnetic field, disappearing for applied fields $\mu_0H\gtrsim\unit[3]{T}$ (Fig.~\ref{fig:susc}B). In Fig.~\ref{fig:susc}B we show the response of the magnetization $M$ to an applied field, observing an abrupt kink at $\mu_0H^*=\unit[2.8]{T}$, as noted by other authors~\cite{takayama_hyperhoneycomb_2015}. This kink occurs at an induced moment of $M^*=\unit[0.31]{\nicefrac{\upmu_{\textrm{B}}}{Ir}}$, far away from the expected saturation magnetization $\sim\unit[1]{\upmu_B}$ for J$_{\textrm{eff}}=\nicefrac{1}{2}$, and does not show any step-like features characteristic of first-order meta-magnetic transitions. Assuming $H^*$ defines a thermodynamic boundary, it is approximately temperature independent below $\unit[25]{K}$.

\begin{figure*}[ht]
    	\includegraphics[width=2\columnwidth]{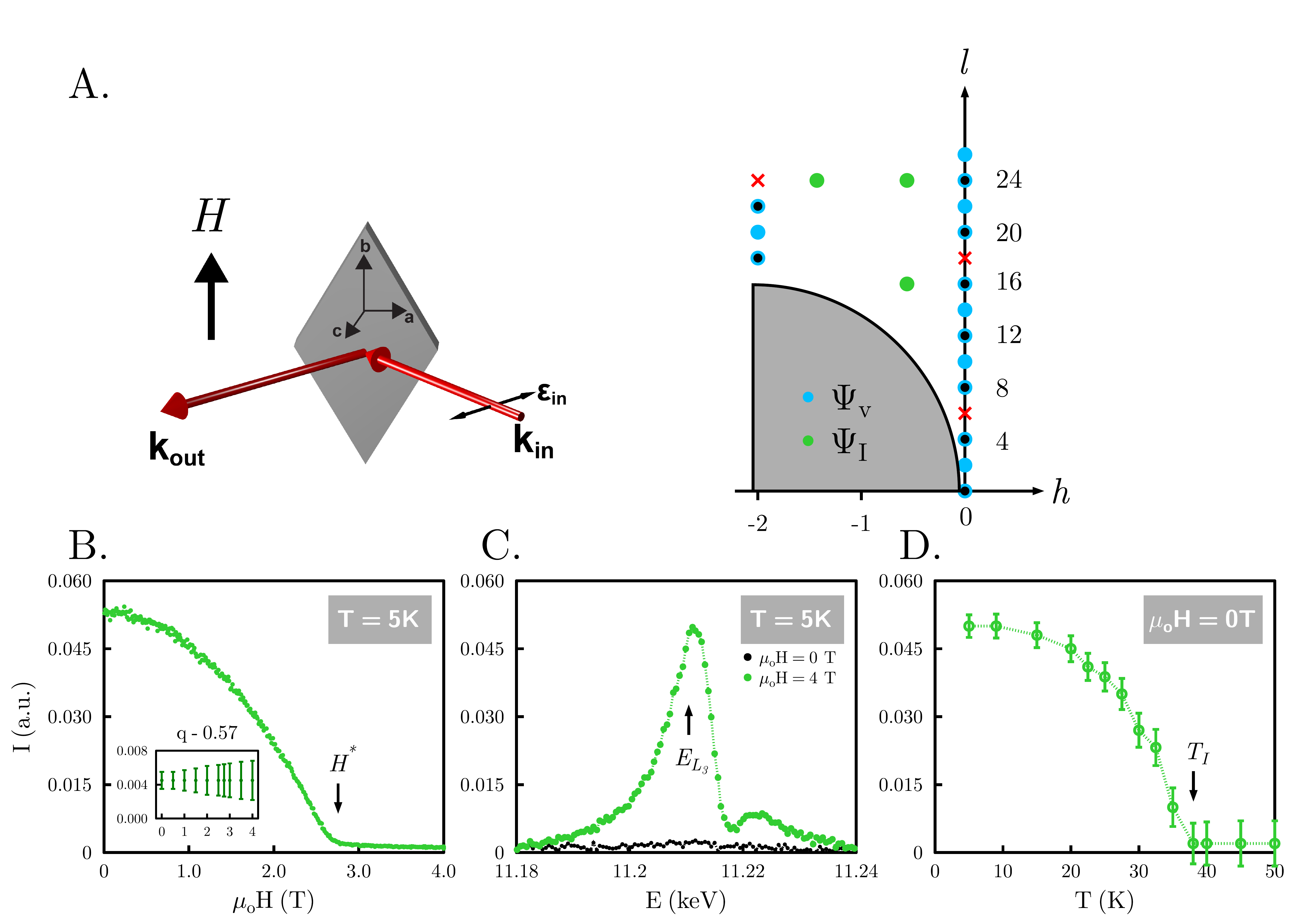}
  	\caption{ {{\bf Fate of the incommensurate order $q=(0.574,0,0)$ under an applied field.}  (A) The scattering geometry used during this experiment showing the polarization of the incoming x-ray beam ($\pi$-polarized) and the direction of the applied magnetic field (along the $\hat{b}$-axis). The right panel displays all the surveyed positions in reciprocal space at $\unit[4]{T}$. Black dots denote structural peaks, while the green and blue dots represent the incommensurate $\Psi_{\textrm{I}}$ and commensurate  $\Psi_\textrm{V}$ magnetic peaks, respectively. The shaded area represents an inaccessible region below the sample horizon. (B) Field dependence of the scattering intensity at $T=\unit[5]{K}$ showing complete suppression at $\mu_0H^*=\unit[2.8]{T}$. The inset shows that the $q$-vector remains constant under an applied magnetic field.} (C) Energy dependence of the scattering intensity at $T=\unit[5]{K}$ for $\mu_0H=\unit[0]{T}$ and $\unit[4]{T}$. (D) Temperature dependence at $\mu_oH=\unit[0]{T}$ showing the onset of the order parameter at $T_I=\unit[38]{K}$.} 
  	\label{fig:RMXS}
	\end{figure*}

To better understand this thermodynamic behavior, we have performed magnetic resonant x-ray scattering (MRXS) at the Ir $L_3$-edge (see Fig.~\ref{fig:RMXS}A and Methods). This diffraction technique is sensitive to spin and orbital orders, allowing the identification of magnetic ground states in reciprocal space. In Fig.~\ref{fig:RMXS}A, we illustrate the real space set-up and the $(h,0,l)$ scattering plane that is experimentally accessible. In Fig.~\ref{fig:RMXS}B, we show reciprocal space scans around $Q=(-0.574,0,16)$ as a function of increasing magnetic field at $\unit[5]{K}$. This $q$-vector corresponds to the incommensurate magnetic order that sets in at $T_I=\unit[38]{K}$. The peak intensity is reduced with increasing field (completely vanishing at $H^*$), without changing its wavevector ${\bf q}$ (Fig.~\ref{fig:RMXS}C inset). In other words, although $\Psi_{\textrm{I}}$ is strongly suppressed in magnitude, its translational symmetry remains rigid.  

\begin{figure*}[ht]
    	\includegraphics[width=2\columnwidth]{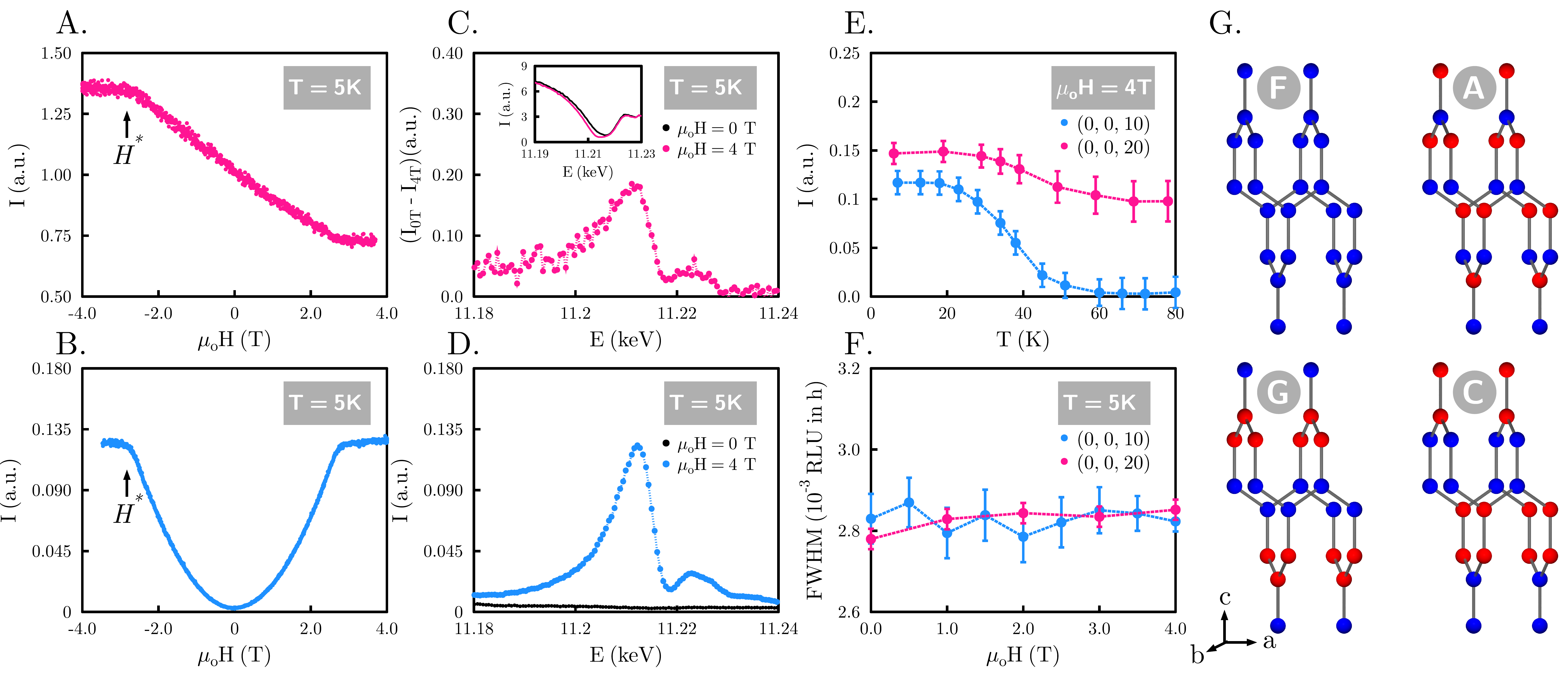}
  	\caption{{\bf Field, energy and temperature dependence of the commensurate order $q=(0,0,0)$.} Field dependence of the scattering intensity taken at $T=\unit[5]{K}$ and $E=\unit[11.215]{keV}$ around: (A) the structurally allowed $h+l=4n$ peaks (e.g. $(0,0,20)$) which show a linear dependence and, (B) the symmetry disallowed peaks $h+l=12n\pm2$ (e.g $(0,0,10)$) which show a quadratic dependence to the applied field. A kink was again observed at $\mu_oH^*=\unit[2.8]{T}$. The energy dependence for the allowed peaks is shown in the inset (C) with a dip at the absoption edge $E=\unit[11.215]{keV}$. The main panel (C) shows the difference between the intensity at $\mu_oH=\unit[0,4]{T}$ which can be attributed to a magnetic contribution. (D) Energy dependence of the magnetic peak $h+l=12n\pm2$ taken at $\mu_oH=\unit[0,4]{T}$. (E) Temperature dependence  of the integrated intensities for the $(0,0,10)$ and $(0,0,20)$ peaks, denoted pink and blue respectively. (F) $(0,0,10)$ and $(0,0,20)$ peaks widths (a lower bound on the correlation length) remain constant under the applied field, suggesting there is no macroscopic phase separation. (G) Possible basis vectors describing the magnetic order of \blio, where $F$ corresponds to ferromagnetic order, $A$ to N\'eel order, $C$ to stripy order and $G$ to zig-zag order.}
  	\label{fig:RMXS_HF}
	\end{figure*}

We now turn our attention to the properties of this material beyond the phase boundary delimited by $H^*$. Our thermodynamic data indicated that a high-field magnetic order was present for $H>H^*$ 
and while no field-induced changes were observed by MRXS at high symmetry (e.g. $h,l=\nicefrac{1}{4},\nicefrac{1}{3},\nicefrac{1}{2}$ etc.) nor any other incommensurate positions, we found intensity changes at certain reciprocal $\it{lattice}$ vectors (blue dots in Fig.~\ref{fig:RMXS}A). In Fig.~\ref{fig:RMXS_HF}A,B we plot the field response of the scattering intensity at two kinds of reciprocal space points, one belonging to structurally allowed $h+l=4n$ peaks and the other to structurally forbidden $h+l=12n\pm2$ peaks. In the former, the response is linear with a negative slope, and shows a kink at $H^*$ (see Fig.~\ref{fig:RMXS_HF}A). In the latter case, we find that the $h+l=12n\pm2$ peaks have a quadratic dependence on $H$, again with a kink at $H^*$ (see Fig.~\ref{fig:RMXS_HF}B). The appearance of structurally forbidden peaks immediately suggests that there is an additional $\mathbf{q}=\mathbf{0}$ broken symmetry. 

The energy dependence of these peaks highlights an important contrast in behavior between structurally allowed and forbidden peaks. The former, as shown in Fig.~\ref{fig:RMXS_HF}C have a dip at the Ir $L_3$-edge due to the resonant absorption of the Ir lattice, while the field-induced change of intensity occurs only near the Ir resonance, suggesting a change in the spin population. Moreover, the structurally forbidden peaks, $h+l=12n\pm2$, are enhanced at the $L_3$-edge in an applied magnetic field (see Fig.~\ref{fig:RMXS_HF}D), suggesting that the $\mathbf{q}=\mathbf{0}$ order is electronic in origin and, as we argue below, most likely magnetic.  

In Fig.~\ref{fig:RMXS_HF}E we illustrate the temperature dependence of the intensity for both sets of peaks at a fixed field of $\unit[4]{T}$ ($H>H^*$). The evolution is closely reminiscent of the temperature dependence of a symmetry-breaking order parameter, turning on at $\sim\unit[50]{K}$. We assign $\Psi_{\textrm{V}}$ as the parameter describing this field induced broken symmetry, which {\it ex concessis} is proportional to the scattering form factor, leading to a quadratic field dependence in the intensity $I_{12n\pm2}$.  The height of a magnetic Bragg peak can grow by virtue of three things: an increase in correlation length, a canting of the moment that enhances scattering cross section, or an increase in the ordered local moment itself $\sigma_i$. In the present case, the intensity of magnetic peaks increases dramatically with $H$, and the peak width remains constant (Fig.~\ref{fig:RMXS_HF}F), ruling out the first scenario (implying macroscopic phase separation is highly unlikely). Moreover, a field-enhanced cross-section would imply a zero-field magnetic order. To see this, note the scattering cross section is proportional to $(\hat{\epsilon}_{out}\times\hat{ \epsilon}_{in})\cdot \hat{m}_i$, where $\hat{\epsilon}_{out (in)}$ denotes the polarization state of the scattered (incident) beam, and $\hat{m}_i$ is a unit vector along the magnetic moment at site $i$. To enhance the cross section, we require a field-induced canting of ordered moments parallel to the term $(\hat{\epsilon}_{out}\times\hat{ \epsilon}_{in})$, and this can only be achieved for one polarization state at a given incidence angle. Considering we are measuring both the $\pi-\sigma$ and $\pi-\pi$ channels, and see no intensity for all $12n\pm2$ peaks at $\mu_oH=\unit[0]{T}$, we can rule out a zero-field phase with continuous canting of the moments by the field. This implies that our observations are most likely explained by an increasing moment size, with long-range quantum correlations turning on $\sim\unit[50]{K}$, which cannot develop a sizable ordered moment at zero field, presumably due to the system's intrinsic magnetic frustration. 
  
This behavior differs from archetypical examples of phase transitions in magnetic field such as spin-flops or incommensurate to commensurate transitions~\cite{chattopadhyay_incommensurate_1986,kiryukhin_soliton_1996,vaknin_commensurate-incommensurate_2004, 1402-4896-71-2-N02, feng_evolution_2013,nagao_experimental_2015,krimmel_incommensurate_2006}. These are usually first order transitions; the former is a transition between an antiferromagnetic state and a spin-polarized state, while the latter will often cause the incommensurate order to soften, shifting toward a commensurate ${\bf q}$ as it is suppressed by the field. In the present case, far from there being a phase transition between one kind of order and another, all broken symmetry states coexist, retaining their intrinsic periodicity as a function of field. The ordered moment, is somehow shared between different ground states, indicating their near degeneracy.

\begin{figure}[ht]
    	\includegraphics[width=9cm]{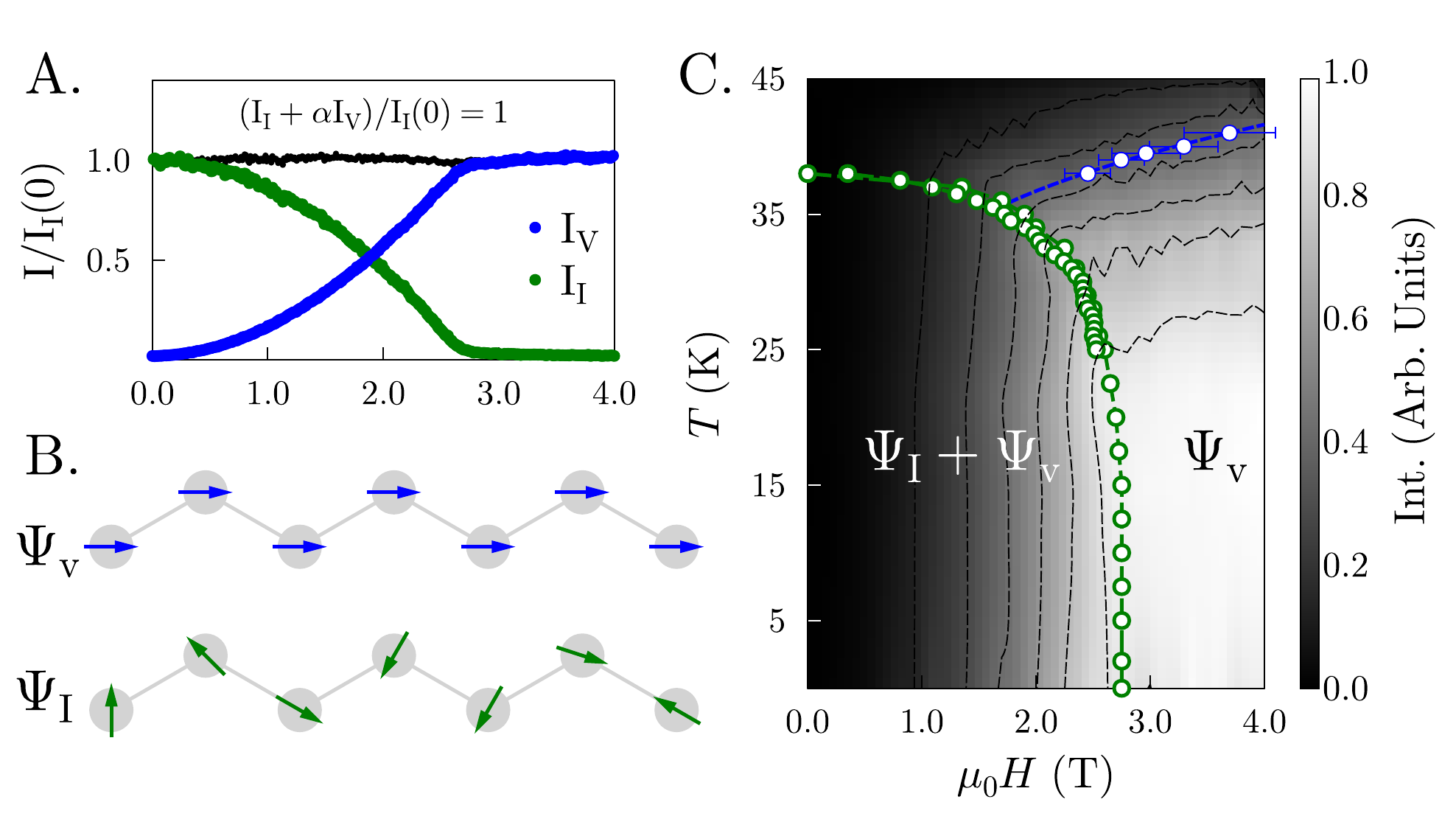}
  	\caption{{\bf T - H phase diagram of \blio\ with field along $\hat{b}$-axis.} (A)  $\Psi_{\textrm{I}}$ and $\Psi_{\textrm{V}}$ obey a simple sum rule which suggests that these are the important order parameters in the system. (B) Schematic representation of the moment direction in the zero field incommensurate (lower, green arrows) and the field induced zig-zag (upper, blue arrows) order projected on the Ir chains (gray) which propagate in the crystallographic $\hat{a}\pm\hat{b}$ directions (see also Fig. \ref{fig:susc}A). Note the propagation vector of the incommensurate order is not along the chain, but along the $\hat{a}$ axis. (C) $T - H$ phase diagram constructed by combining the normalized scattering intensity of the $h+l=12n\pm2$ peaks (gray scale contours), with the $H^*$ and $T_I$ extracted from magnetization and heat capacity measurements (green dots), and a cusp observed in the magnetic field dependence of $C_{ac}$ (blue dots) (see Supplementary Information). Note the order parameter $\Psi_{\textrm{V}}$ in principle grows as the structure factor and so $\Psi_{\textrm{V}}\propto I_V^2$.} 
  	\label{fig:phasedi}
	\end{figure}	

{\bf Discussion.} In Fig.~\ref{fig:RMXS_HF}G we show each of the symmetry distinct ${\bf q}={\bf 0}$ broken symmetry states, denoted using the conventional nomenclature. The states that are consistent with the primary features of the data are two-fold: (i) $G$-type (zig-zag) broken symmetry (explaining the appearance of the $12n\pm 2$ peaks) and (ii) $F$-type order (explaining the linear dependence of the the allowed structural peaks). Importantly, these represent symmetries that are broken by the applied magnetic field itself. First, an induced moment along the applied magnetic field on each ion would transform as an $F$-like object with moments along $\hat{b}$ (denoted $F_b$). Second, the effective local susceptibility tensor (including orbital $g$-factor as well as correlation effects) is anisotropic (see reference~\onlinecite{modic_lio} for a discussion of \glio) and it can be decomposed into components parallel and perpendicular to the two honeycomb chain directions, $\hat{a}\pm\hat{b}$. This implies that the moments will in general cant equally and oppositely along the $\hat{a}\pm\hat{b}$ directions, leading to $G$-like broken symmetry, with moments staggered along $\hat{a}$ (denoted $G_a$). It can be shown (see Supplementary Information) that $F_b$ and $G_a$ not only belong to the same irreducible representation as $\Psi_{\textrm{I}}$, allowing $\Psi_{\textrm{V}}$ to coexist on symmetry grounds, but are in fact energetically favored over other broken symmetry states. Our observations are completely consistent with Landau theory of second order phase transitions - the combined effect of magnetic field and crystal symmetry is to act as a `field' for the commensurate order, so that the observed zig-zag pattern is linearly coupled to magnetic field (leading to an intensity $I_V$ that is quadratic, see Fig.~\ref{fig:RMXS_HF}B). This also explains why there is no thermodynamic anomaly on cooling at $H>H^*$ (Fig.~\ref{fig:susc}C):  the symmetry breaking associated with $\Psi_{\textrm{V}}$ is already broken by the applied $\it{H}$. 

However, a simple linear response to the applied field does not explain the temperature dependence of $\Psi_{\textrm{V}}$ (Fig.~\ref{fig:RMXS_HF}E), which evolves like a thermodynamic order parameter. Nor does it explain the observation that the moment lost from $\Psi_{\textrm{I}}$ appears to be entirely transferred to $\Psi_{\textrm{V}}$, which is strongly suggested by a simple sum rule of their respective intensities (Fig.~\ref{fig:phasedi}A). Indeed, if we assume that the ordered moment of $\Psi_{\textrm{V}}$ is canted entirely along the diagonal bond axes $\hat{a}\pm\hat{b}$ (which would be equivalent to the `zig-zag' order seen in \anio\,and RuCl$_3$, we find a quantitative equivalence between the ordered moment of $\Psi_{\textrm{V}}$ and $\Psi_{\textrm{I}}$. Noting at the kink field the magnetic moment is $\sim\unit[0.31]{\nicefrac{\upmu_{\textrm{B}}}{Ir}}$, this suggests that the moment along the chains is $0.31/{\cos(0.2\pi)}$, where $0.2\pi$ is around half the angle between the diagonal chains. This yields $\sim\unit[0.4]{\upmu_{\textrm{B}}}$, accounting for most of the moment in $\Psi_{\textrm{I}}$, measured independently to be $\sim\unit[0.47]{\upmu_{\textrm{B}}}$ at zero field~\cite{biffin_unconventional_2014}. 

Nevertheless, the heat capacity (Fig.~\ref{fig:susc}B) shows no singular anomaly expected of a thermodynamic phase transition for $\mu_0H>\unit[2.8]{T}$, but instead shows a broad break in slope as the temperature is lowered. This break in slope is much clearer in the field-dependent heat capacity at constant temperature (see Supplementary information), where it appears as a broad crossover between the low field and high field regimes. Plotting the location of this peak on a phase diagram, we observe immediately that it maps directly onto a constant contour line of the I$_{12n\pm2}$ amplitude, the experimental measure of $\Psi_{\textrm{V}}$ itself. In other words, the applied magnetic field, induces a thermodynamically stable `zig-zag' ordered phase, but without a spontaneous symmetry-breaking phase transition or its associated singularities. 

 In $\alpha$ type structures, the low temperature ground state is found to be incommensurate~\cite{williams_incommensurate_2016} in the Li-based compounds but zig-zag in the Na-based compounds. The former has been linked to strong Kitaev exchange in several independent theoretical works \cite{kimchi_unified_2015, lee_theory_2015}, whereas the latter does not directly need Kitaev exchange at all~\cite{choi_spin_2012}, leaving the question open as to whether a single Hamiltonian captures the relevant physics for all of these compounds.  Our results hint that there exists a universal theoretical framework with which to understand the magnetism of the `harmonic' honeycomb iridates, with Kitaev exchange as the dominant energy scale. In fact, given a simple model of Kitaev and Heisenberg exchange terms in the effective Hamiltonian, it can be shown that the transition to $\Psi_{\textrm{V}}$ at $H^*$ directly implies that the Kitaev term dominates (See Supplementary Information Sec.\ref{sec:KHest}). In addition, this work exposes the presence of a quantum critical point at $H^*$ in a Kitaev exchange-dominant system~\cite{kimchi_unified_2015}, providing a unique opportunity to study the physics of Kitaev interactions in a purely quantum ($T=\unit[0]{K}$) regime~\cite{janssen_honeycomb-lattice_2016}.

\blio\, has a single symmetry distinct magnetic site, and this makes the behavior of $\Psi_{\textrm{V}}$ particularly dichotomous. On the one hand it behaves much like a ferromagnet in an applied field, where the field breaks the same symmetry as the order parameter such that no true phase transition occurs. In this view, our observations are completely consistent with the Landau theory of second order phase transitions. On the other hand, while a ferromagnet has a finite order parameter at zero field, in the present case $\Psi_{\textrm{V}}$ only exists if a field is present. In this sense, the applied magnetic field is acting on a {\it vestigial} order, taking moment away from the true ground state $\Psi_{\textrm{I}}$ and transferring it to another $\Psi_{\textrm{V}}$, without ever causing a thermodynamic phase transition. Intriguingly, the symmetry of $\Psi_{\textrm{V}}$ is the exact analogue of the the ground state of related Mott-Kitaev systems RuCl$_3$ and \anio. The closeness of zig-zag and incommensurate states is a striking experimental signature of the intrinsic frustration of these systems. Indeed, it is likely that there are many other ordered states that are nearly degenerate, but only form when a field compatible with their symmetry is present. In addition to placing severe constraints on the microscopic Hamiltonian describing these systems, these results experimentally demonstrate that there is a unified description of the Mott-Kitaev honeycomb family~\cite{kimchi_unified_2015}, and more broadly, that these materials are playgrounds for the study of exotic phases in highly frustrated quantum magnets.

\section{Supplementary Information}
\subsection{Sample Synthesis and crystal structure}
\begin{figure}[ht]
    	{\includegraphics[width=9cm]{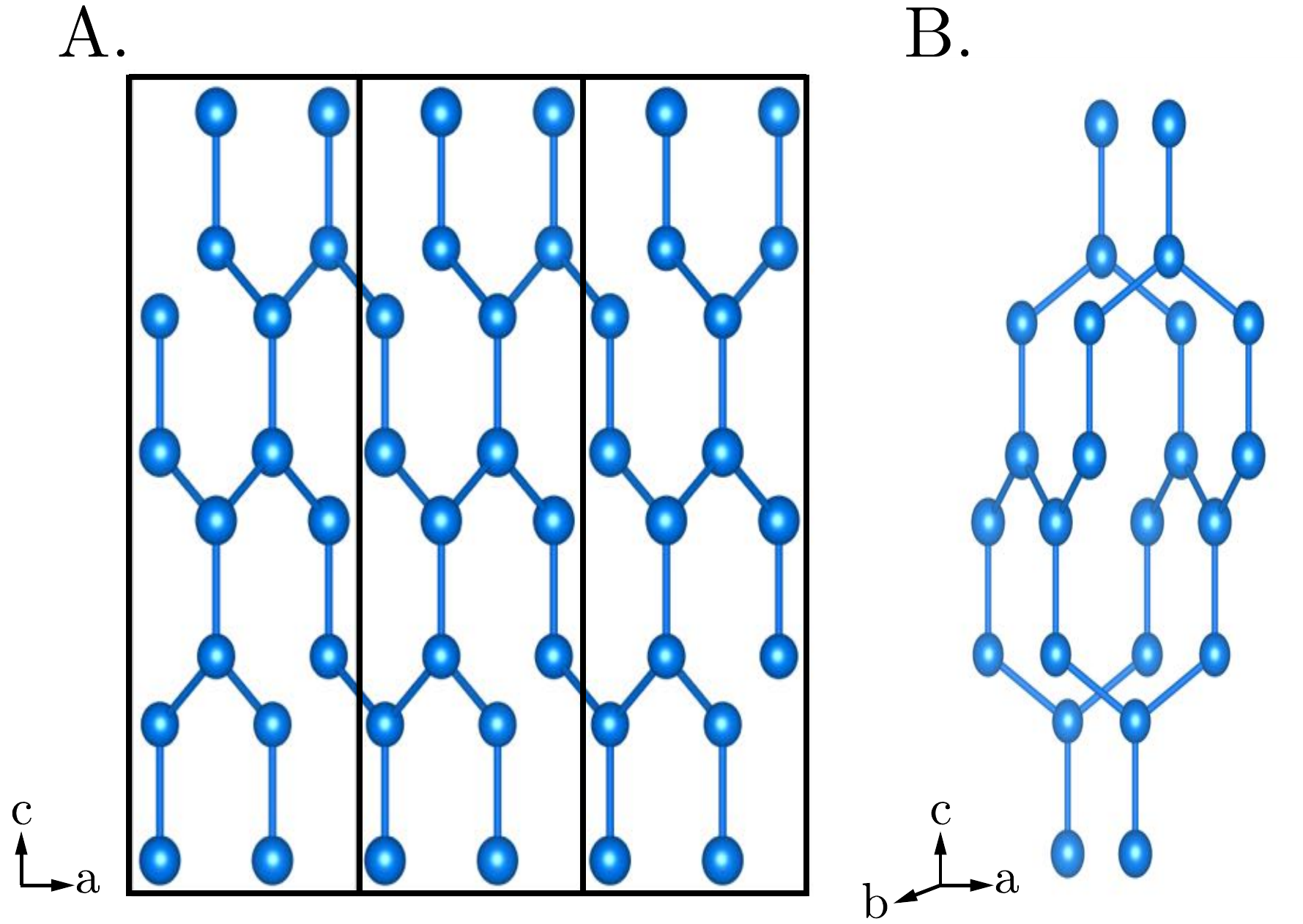}
  	\caption{ {\bf \blio\ crystal structure.} (A) Projection in the $\it{ac}-$plane. (B) three-dimensional view of a unit cell. The Ir atoms (blue dots) form zigzag chains stacked along the $\hat{c}$ and directed alternatingly along $\hat{a}\pm \hat{b}$. }
  	\label{fig:crys}}
	\end{figure}

Single crystals of \blio\ were synthesized using a vapor transport technique. Ir (99.9\% purity, BASF)  and Li$_2$CO$_3$ (99.999 \% purity, Alfa-Aesar) powders were grounded and pelletized in the molar ratio of 1:1.05. The pellet was placed on an alumina crucible and reacted at 1,050 \textdegree C for 12hrs, and then cooled down to 850 \textdegree C at 2 \textdegree C/hr to yield a powder sample containing single crystals which are clearly faceted and around 105$\times$150$\times$300 $\upmu$m$^3$ in size. Room temperature powder and single crystal x-ray diffraction indicated that the high quality crystals were \blio\ with an orthorhombic crystal structure and selection rules consistent with the {\it Fddd} space group (Figure \ref{fig:crys}).\\

\begin{table}[!hb]
\begin{center}
{\renewcommand{\arraystretch}{1.5}
\begin{tabularx}{0.45\textwidth}{ >{\setlength\hsize{1\hsize}\centering}c|>{\setlength\hsize{1\hsize}\centering}X >{\setlength\hsize{1\hsize}\centering}X >{\setlength\hsize{1\hsize}\centering}X }
\hline \hline\
{\bf Z} & 16 \tabularnewline
{\bf Space Group:}& {\it Fddd}\tabularnewline 
{\bm $a$,$b$,$c$ (\AA):} & 5.910(1) & 8.462(2) & 17.857(6) \tabularnewline
{\bm $\alpha$,$\beta$,$\gamma$ (\textdegree) :} & 90\textdegree & 90\textdegree & 90\textdegree\tabularnewline
{\bf Volume (\AA$^3$):} & 893.0(5) & & \tabularnewline
\hline \hline 
\end{tabularx}}
\caption{Structural Parameters of \blio\ at \unit[300]{K}.}
\end{center}
\end{table}

This 3D structure is locally identical to the 2D honeycomb lattice, \alio\, in which each IrO$_6$ octahedra shares an edge with three neighbors. The difference arises due to a bonding degeneracy between the edge-sharing octahedra, which results in a three-dimensional network of $Ir$ moments \cite{modic_lio}. \\
\subsection{Thermodynamic Properties}
\subsubsection{General Properties}
In this section, we provide further information about the thermodynamic properties of single crystal \blio. Magnetic susceptibility measurements were performed in a 7T Cryogenic S700X and a Quantum Design MPMS3 SQUID. Specific heat measurements were conducted on a 16T Cryogenic CFMS using the {\it ac} calorimetry method, which detects oscillations on the sample's temperature in response to an oscillating heat power \cite{kohama_ac_2010, minakov_thin-film_2005}. For this, a sample is placed over six thermocouples connected in series under a free standing silicon nitride membrane $\sim$ 1$\upmu$m thick. An {\it ac} current, with frequency $\omega$, is driven through an adjacent resistive heater, resulting on an oscillating power, $P_{ac}=\nicefrac{1}{2}I_o^2\cdot R\cdot(1+cos(2\omega t))$. The resulting sample's temperature ($V_{ac}$) oscillates at frequency $2\omega$ and it is used to calculate the ac heat capacity:
\begin{equation}
C_{ac}=\frac{K(T) \cdot P_{ac}}{\omega \cdot V_{ac}}
\end{equation}
These measurements are performed in a low pressure He-4 gas environment ($\sim\unit[10]{mbar}$). The optimal frequency range used was $\unit[20]{Hz}$, necessary to ensure that the thermal link through the membrane and the gas can be ignored, and  that the sample is heat homogeneously \cite{riou_determination_2004}. 
   
\begin{figure}[ht]
    	{\includegraphics[width=9cm]{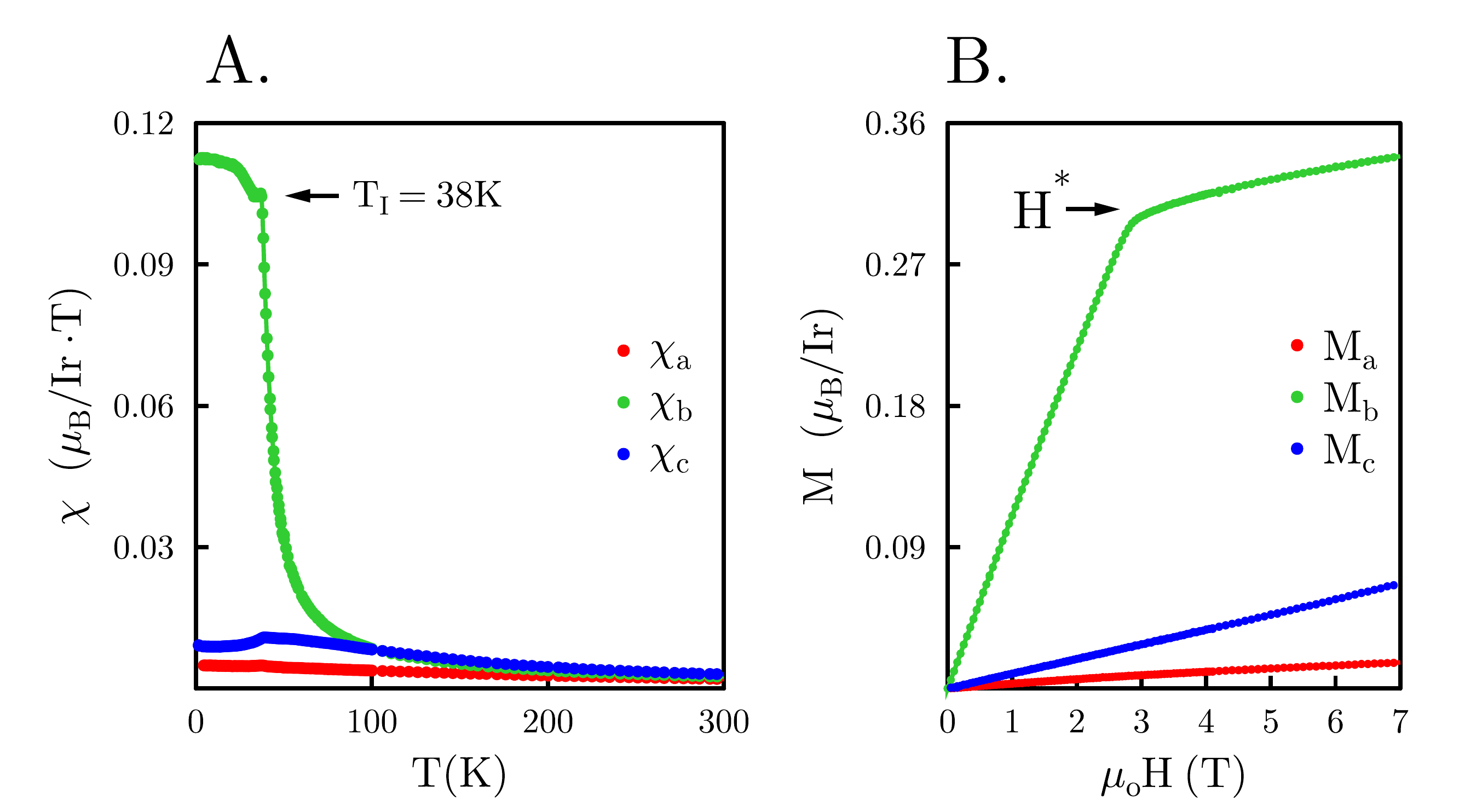}
  	\caption{ {\bf Magnetic properties of a single crystal of \blio\ along the three principal crystallographic orientations.} (A) Temperature dependence of the magnetic susceptibility at 1T. This material orders into an incommensurate magnetic structure at $T=\unit[38]{K}$. Notice the highly anisotropic behavior of this $J_{\textrm{eff}}=\nicefrac{1}{2}$ system with  $\chi_a$:$\chi_b$:$\chi_c$ $\approx$ $1 : 40 : 10$  (B) Field dependence of magnetization at $\unit[5]{K}$. The response of the $\hat{a}$ and $\hat{c}-$axis is linear with field up to 7T. Meanwhile the $\hat{b}-$axis responds linearly up to a kink field $H^*$.}
  	\label{fig:all_axes}}
	\end{figure}

Our specific heat and susceptibility measurements showed a clear anomaly at T$_I=\unit[38]{K}$, in good agreement with what has been previously reported (a complex, incommensurate magnetic ground state with non-coplanar and counter-rotating $Ir$ moments that sets in at $T_I$) \cite{biffin_unconventional_2014}. Figure \ref{fig:all_axes} shows the magnetic properties of \blio\ with field applied along the principal axes. The response to an applied magnetic field at low temperatures is very anisotropic with $\chi_a$:$\chi_b$:$\chi_c$ $\approx$ $1 : 40 : 10$ (see Figure \ref{fig:all_axes}). A linear Curie-Weiss behavior (eq. 2) was observed for $T>\unit[150]{K}$, with overall effective moment $\mu_{eff}=\unit[1.81]{\nicefrac{\upmu_B}{Ir}}$, and Curie-Weiss temperature $\Theta_{CW}=\unit[-30]{K}$.
\begin{equation}
\chi=\frac{\mu_oN_A\mu^2_{eff}\mu^2_B}{3k_B(T-\Theta_{CW})}
\end{equation}

\begin{table}
\begin{center}
{\setlength{\extrarowheight}{2.5pt}
\begin{tabularx}{0.45\textwidth}{ >{\setlength\hsize{1\hsize}\centering}X >{\setlength\hsize{1\hsize}\centering}X >{\setlength\hsize{1\hsize}\centering}X >{\setlength\hsize{1\hsize}\centering}X }
\hline \hline 
\multicolumn{4}{c}{Fitting at 1T along principal axes.}\tabularnewline
\hline
  & {\bm $\hat{a}$} & {\bm $\hat{b}$} & {\bm $\hat{c}$} \tabularnewline
\hline
{\bm $\upmu_{eff}(\nicefrac{\upmu_B}{Ir})$ } & $1.86\pm0.1$ & $1.76\pm0.1$ & $1.99\pm0.1$ \tabularnewline
{\bm $\uptheta_{CW}(K)$} & $-94\pm5$ & $18\pm5$ & $0\pm5$ \tabularnewline
\\
\multicolumn{4}{c}{Fitting for multiple fields along $\hat{b}-$axis.}\tabularnewline
\hline 
  & {\bf 1.0T} & {\bf 2.0T} & {\bf 4.0T} \tabularnewline
\hline
{\bm $\upmu_{eff}(\nicefrac{\upmu_B}{Ir})$ } & $1.76\pm0.1$  & $1.73\pm0.1$ & $1.76\pm0.1$ \tabularnewline
{\bm $\uptheta_{CW}(K)$} & $18\pm5$  & $20\pm5$ & $18\pm5$  \tabularnewline
\hline \hline
\end{tabularx}}
\end{center}
\caption{Currie-Weiss Parameters of \blio\ fitted for $T>\unit[150]{K}$.}
\label{Currie-Weiss}
\end{table}
The effective moment is very close to the expected value $\upmu_{\textrm{eff}}=\unit[\sqrt{3}]{\nicefrac{\upmu_{\textrm{B}}}{Ir}} \sim\unit[1.73]{\nicefrac{\upmu_{\textrm{B}}}{Ir}}$ for the ideal $J_{\textrm{eff}}=\nicefrac{1}{2}$ moment. The Curie-Weiss temperature is close to the transition temperature for the incommensurate order, $\Theta_{_\textrm{CW}}$ $\sim T_I$, as is the case for unfrustrated magnets. However, this value is the result of cancellations between the ferromagnetic and antiferromagnetic interactions, so the frustration parameter, $f=\nicefrac{\Theta_{\textrm{CW}}}{T_I}$, is not a good indication of the degree of frustration in this material. This can be seen in table \ref{Currie-Weiss}, when we fit each axis independently to a Currie-Weiss model. It implies that the interactions along $\hat{b}-$axis are weakly ferromagnetic while the $\hat{a}-$axis interactions have a bigger antiferromagnetic nature. We should also point out that this results vary drastically depending in the temperature range used for the fitting. We have decided to fit our data between $150-\unit[300]{K}$, since below that, it deviates very strongly from a linear behavior and thus, a simple Curie-Weiss model is not applicable.

Figure \ref{fig:all_axes}B shows the field dependence of magnetization at $T=\unit[5]{K}$.  Note that there is no sign of hysteresis with field applied along any direction. There is also no difference between field-cooled and zero field-cooled curves at high fields. However, M$_a$ and M$_c$ respond linearly to an increasing magnetic field up to \unit[7]{T} while M$_b$ has a linear response up to a kink field $\upmu_o$H$^*$ $\sim$ \unit[2.8]{T}, followed by a gradual increase. The magnetization at H$^*$ is $\sim\unit[0.31]{\nicefrac{\upmu_{\textrm{B}}}{Ir}}$ well below the expected $\upmu_\textrm{B}$ for a fully polarized $J_{\textrm{eff}}=\nicefrac{1}{2}$ isospin. 
  	
\begin{figure*}[ht]
    	{\includegraphics[width=16cm]{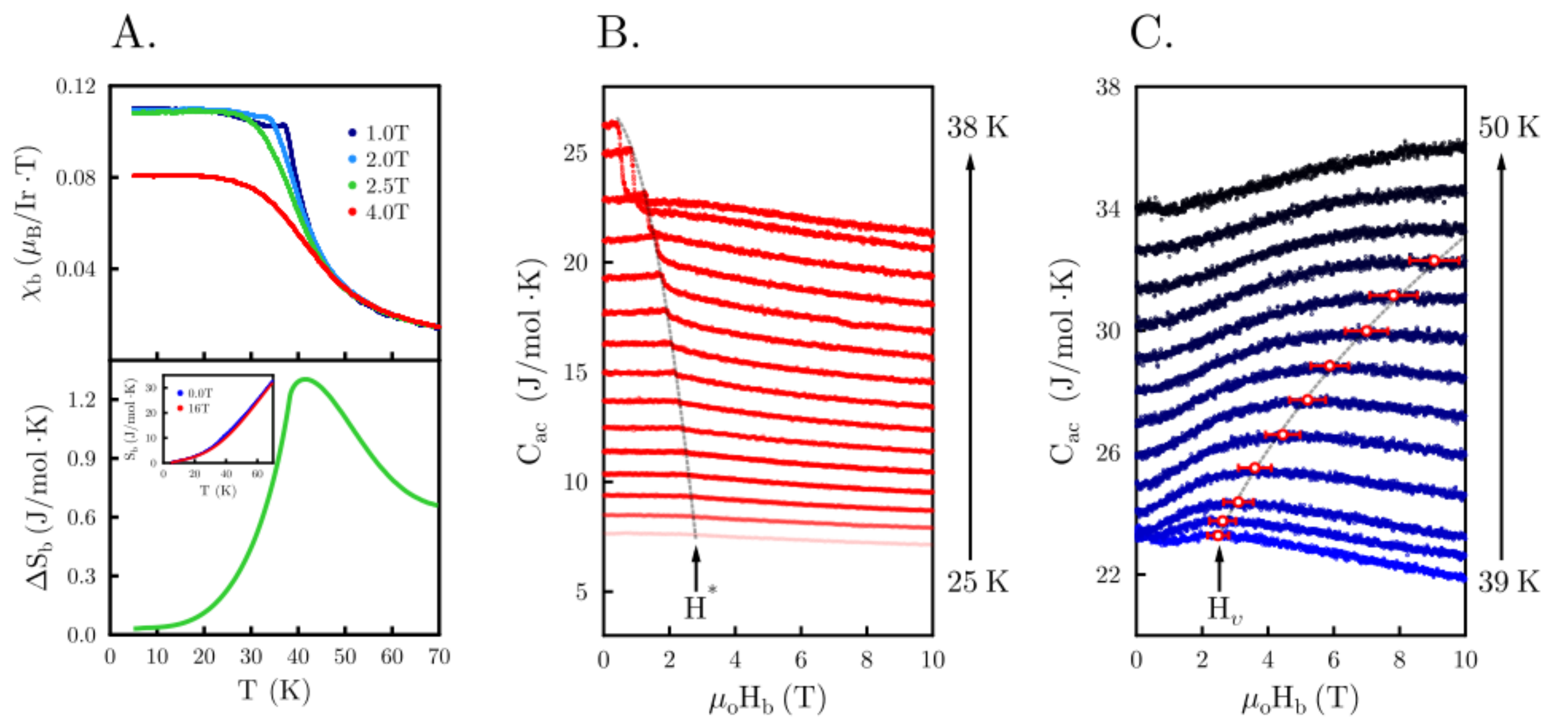}
  	\caption{ {\bf Thermodynamic properties of\blio\ with field applied along the $\hat{b}-$axis.} (A) Susceptibility as a function of temperature for $\mu_oH=$ 1, 2, 2.5 $\&$ \unit[4]{T}. The lower panel shows the total entropy at $\mu_oH=$ 0 \& \unit[16]{T}. The inset shows the entropy change which represents $\sim22.5\% R\ln2$ . (B) Field dependence of the heat capacity for $T<T_I$ and (C) $T>T_I$. A broad hump in the data indicates a crossover into a field-induced phase. A dotted line has been added as a guide to the eye with error bars, of width $2\epsilon$, determined self consistently using $\vert y(H_{max}) - y(H_{max}-\epsilon)\vert = \Delta y$, where $\Delta y$ is the RMS noise in the measurement.}
  	\label{fig:b_axis}}
\end{figure*}  
\subsubsection{Magnetic response with field applied along the $\hat{b}-$axis}
Let us turn our attention to the $\hat{b}-$axis properties of \blio. The edge-sharing IrO$_6$ octrahedra preserve the essential physics of the Kitaev model where interfering Ir$-$O$_2-$Ir exchange paths give rise to orthogonal component of spin\cite{jackeli_mott_2009}. 
\begin{equation}
H_K=K^{\alpha}\sum_{\langle ij\rangle}^{\alpha \epsilon (x,y,z)}{S_i^{\alpha}S_j^{\alpha}}
\end{equation}
This Hamiltonian can be relabeled using the crystallographic directions of the 3D orthorhombic honeycomb iridates \cite{modic_lio}:
\begin{align*}
H_K= & -K^c\sum_{\langle ij\rangle\epsilon \hat{b}_{}}{S_i^{\hat{b}}S_j^{\hat{b}}}\\
& -K^h\sum_{\langle ij\rangle\epsilon (\hat{a}+\hat{c})_{}}{S_i^{\hat{a}+\hat{c}}S_j^{\hat{a}+\hat{c}}}\\
& -K^h\sum_{\langle ij\rangle\epsilon (\hat{a}-\hat{c})_{}}{S_i^{\hat{a}-\hat{c}}S_j^{\hat{a}-\hat{c}}}
\end{align*}
where $S^{\hat{b}}$ and $S^{\hat{a}\pm \hat{c}}=\nicefrac{(S^{\hat{a}}\pm S^{\hat{c}})}{\sqrt{2}}$ are the spin operators in a set of orthogonal directions, with $\hat{a}, \hat{b}, \hat{c}$ being unit vectors along the orthorhombic crystal axis. Each $\langle ij\rangle$ bond is defined by the axis perpendicular to its Ir$-$O$_2-$Ir plane which lies along one of the directions \{($\hat{a} + \hat{c}$), ($\hat{a}-\hat{c}$),$\hat{b}$\}. All the nearest neighbor Ir$-$Ir bonds can be divided into three classes, one for each component of spin: the $\hat{b}$ component from the $\hat{c}-$axis bonds, and the $\hat{a}\pm\hat{c}$ components from the $h$ bonds defining each honeycomb plane. The exchange couplings K$^h$ are constrained by the symmetry of the space group to be the same on the ($\hat{a}\pm\hat{c}$) bonds, but K$^c$, the coeficient of S$^b$ coupling, is symmetry-distinct from K$^h$.  Therefore, the $\hat{b}-$axis is the only crystallographic axis that coincides with an exchange direction in the Kitaev Hamiltonian, making it magnetically special. 

Figure \ref{fig:b_axis}A shows the magnetic susceptibility as a function of temperature taken at 1, 2, 2.5 $\&$ \unit[4]{T}. For $H<H^*$, the susceptibility is constant, with a maximum value of $\sim\unit[0.11]{\nicefrac{\upmu_\textrm{B}}{Ir\cdot T}}$. However, for $H>H^*$ the magnetic susceptibility monotonically decreases, a behavior typical of ferromagnets. Fitting these data to a Curie-Weiss model yielded effective moments $\sim\unit[1.73]{\nicefrac{\upmu_\textrm{B}}{Ir}}$ and positive Weiss temperatures indicative of ferromagnetic interactions (Table \ref{Currie-Weiss}). Also notice that $\chi$(T) at $\mu_oH=\unit[4]{T}$ resembles what is expected for a ferromagnetic order parameter under an applied field \cite{takayama_hyperhoneycomb_2015}. 

The heat capacity data at $\mu_oH=\unit[0]{T}$ and $\unit[16]{T}$ were used to calculate the magnetic entropy change (Figure \ref{fig:b_axis}A inset) associated with the transition. At $T_I$, the value of $\Delta S_m$ $\sim\unit[1.30]{\nicefrac{J}{mol\cdot K}}$ which represents $\sim$ 22.5\% of  $R \cdot\ln2$, the value expected for the magnetic entropy of a $J=\nicefrac{1}{2}$ moment. 

Figure \ref{fig:b_axis}B,C illustrate the heat capacity of \blio\ as a function of applied field for $T<T_I$ (B) and $T>T_I$ (C). Figure \ref{fig:b_axis}B illustrates that near $T_I$ the incommensurate phase transition can be easily distinguished as a sharp break on $C_{ac}(H)$ slope, but as the temperature is lowered, there is almost no difference as one crosses $H^*$, indicating that there is a small entropy change at low temperatures at $H^*$ and that the phase boundary for the incommensurate phase is strongly vertical (consistent with our $M(H)$ data. See main text). However, at $T>T_I$, we observe a broad hump in the field dependent heat capacity, the local maximum of which is marked by $H_v$ in Figure \ref{fig:b_axis}C. This hump, as well as the reduced ordered moment observed in magnetization measurements and the spreading of the magnetic entropy well above $T_I$, indicated that the system is highly frustrated and let us to conclude that a new magnetic order was  induced by field. We constructed a $T-H$ phase diagram that includes $H^*$ (from M(H) measurements) $T_I$ (from $\chi(T)$ measurements) and $H_v$ (from $C_{ac}(H)$), and delineates two phases: the low-temperature, low-field INC order, and the field-stabilized ZZ+ order (described in the main text). 
\subsection{Resonant X-Ray Diffraction Experiment}	
To determine the nature of this new order phase, we performed resonant x-ray scattering experiments at the Ir$-$L$_3$ edge (E = \unit[11.215]{keV}) using a Huber $\Psi$-diffractometer located in beamline 6ID-C at the Advanced Photon Source - Argonne National Laboratory. The sample used was a clearly faceted single crystal of \blio\ about $100\times 150\times 150$ $\upmu$m$^3$ which was glued onto a copper mount using very low quantities of Stycast 1266 epoxy and, its quality and alignment were checked using the x-ray micro-diffraction facility at the Advanced Light Source - Lawrence Berkeley National Laboratory (beamline 12.3.2). The diffraction experiments were carried out in a reflection geometry, using a $\pi$-polarized incident beam $\sim$ $150\times 150$ $\upmu$m$^2$, with the crystal mounted so that the $\hat{b}-$axis was parallel to the applied magnetic field. A split-coil magnet, mounted on the cold finger of a closed-cycle $He$ cryostat with three 60$^{\circ}$ $Be$ windows, provided up to \unit[4]{T} of continuous magnetic field and sample temperature as low as that of liquid He. A horizontal scattering geometry allowed us to measure both $\pi$-$\pi$ and $\pi$-$\sigma$ channels and the scattering data was collected using a photodiode point detector. Due to the size of the split gap, we were only able to access $\pm$ 3.4$^{\circ}$ in the vertical direction while keeping the magnetic field parallel to the $\hat{b}-$axis. Therefore, we were only allowed to survey the $(h, 0, l)$ plane.    
 \begin{figure}[hb]
    	{\includegraphics[width=9cm]{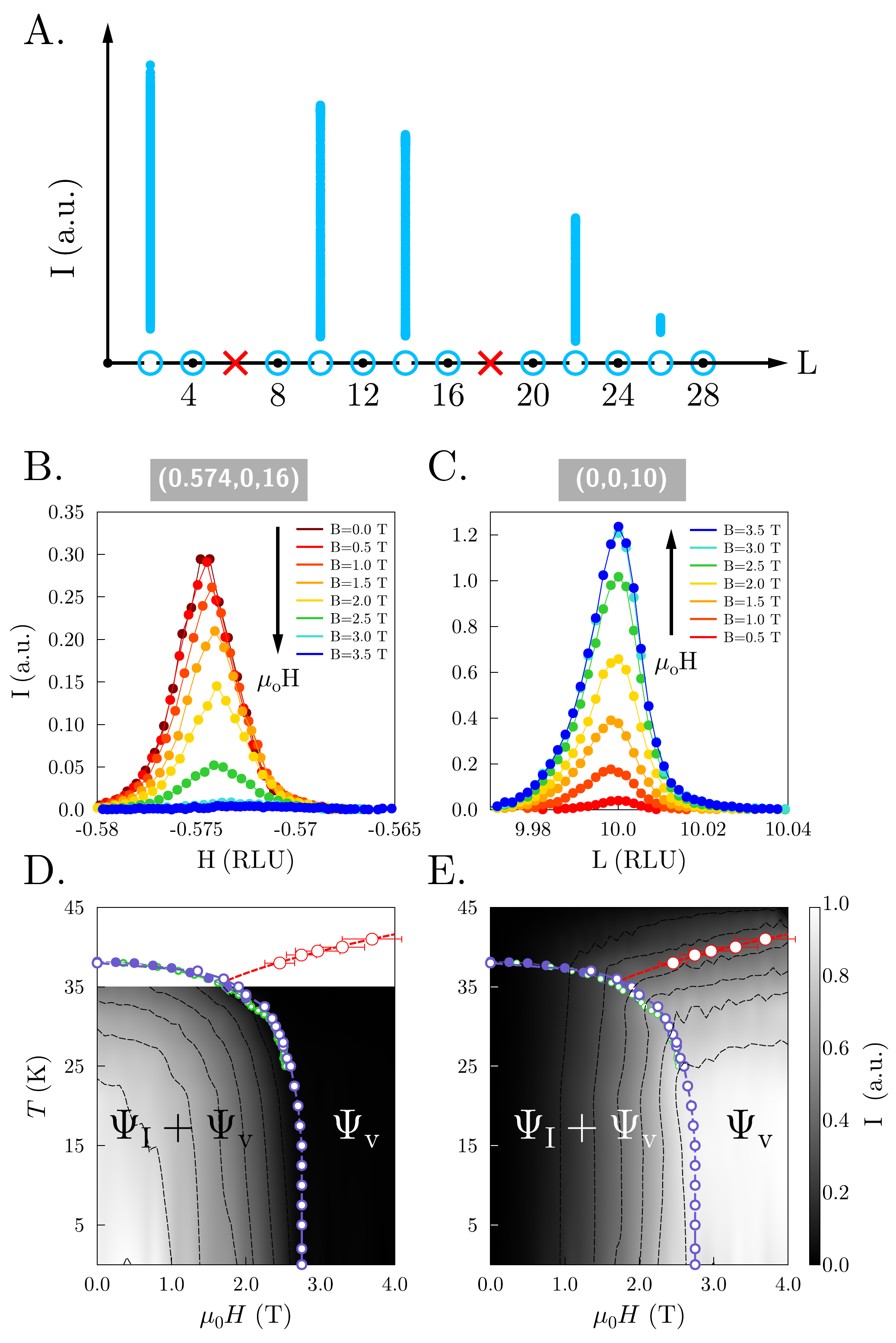}
  	\caption{ {\bf Fate of the spiral order under an applied magnetic field.} The purple line corresponds to the thermodynamic data. Notice that the intensity is suppresed with field and the incommensurate order is completely destroyed for $H>H^*$.}
  	\label{fig:spiral}}
\end{figure} 
The first part of the experiment focused on the field dependence of the incommensurate magnetic order. For this, we studied the behavior of $(0,0,16)+q$, $(0,0,24)-q$ and $(-2,0,24)+q$ with $q=(0.57(4),0,0)$ and were able to show that $\mu_oH^*\sim\unit[2.8]{T}$ completely suppresses the incommensurate order as illustrated by the Figure \ref{fig:spiral}.   (More details on main text).

\begin{figure*}[ht]
    	{\includegraphics[width=15cm]{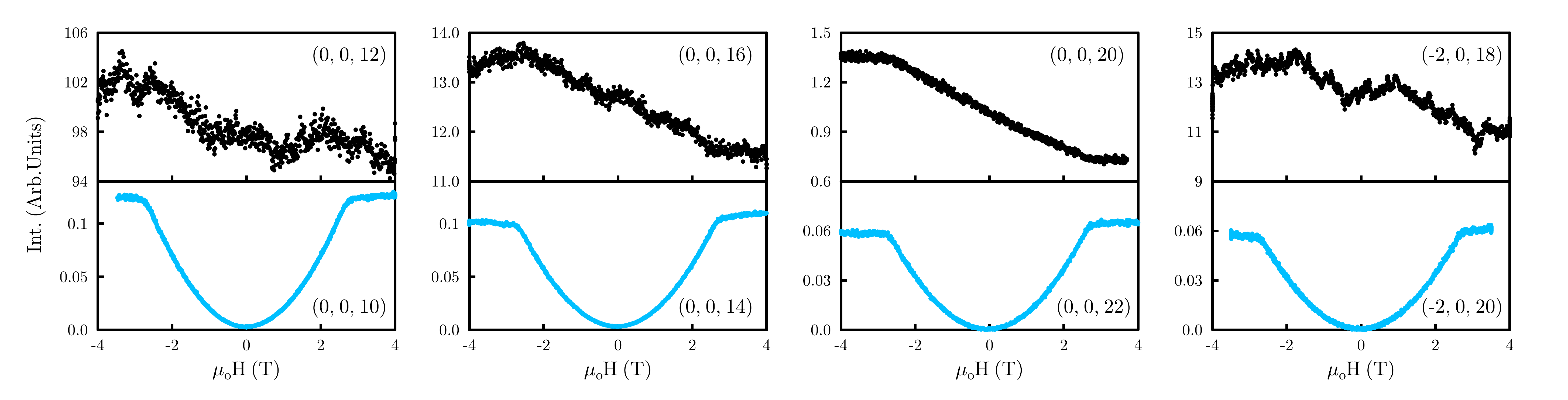}
  	\caption{{\bf Field sweeps at the structural peaks $(0, 0, 4n)$ and the magnetic peaks $(0, 0, 12n\pm2)$.}}
  	\label{fig:field}}
	\end{figure*}

Since our thermodynamic measurements indicated that a new order could be found beyond $H^*$ (Figure \ref{fig:b_axis}D), we continued surveying reciprocal space with $H>H^*$, starting with high symmetry position (\nicefrac{1}{4}, \nicefrac{1}{3}, \nicefrac{1}{2} etc.) and found that a new commensurate magnetic order with $q=(0,0,0)$ is enhanced by the applied field. Figure \ref{fig:field} shows the field dependence of some selected $(h,0,l)$ peaks: the structurally allowed $h+l=4n$ (top panel) and the symmetry disallowed $h+l=12n\pm2$ (bottom panel). Notice that for the $h+l=4n$ peaks, the intensity changes linearly with field, while the $h+l=12n\pm2$ peaks, the intensity is quadratic in field, which indicates that the order parameter couples linearly with $H$, at least for $H<H^*$ (see main text for details). 
We now consider the possible symmetry allowed states for $\Psi_V$, the commesurate order parameter. There is one magnetic Ir$^{4+}$ site, and four such ions in the \blio\,primitive unit cell, connected by the symmetry of the $Fddd$ space group. As discussed in previous work \cite{biffin_noncoplanar_2014}, any commensurate magnetic order can be represented by a four dimensional vector of the relative phases of the Fourier components at the four magnetic sites in the primitive unit cell. Table \ref{position} lists the positions of these ions in the orthorhombic unit cell, with $z=0.70845(7)$, and the possible basis vectors obtained using the {\it BasIReps} tool in FULLPROF and assuming a magnetic structure with propagation vector $q=(0,0,0)$. According to the language of Ref. \onlinecite{kimchi_unified_2015}, $F$ corresponds to ferromagnetic order, $A$ corresponds to N\'eel order, $C$ to stripy order and $G$ to zig-zag order. The real space configuration of magnetic ions and their relative phases are shown in Figure 3G of main text. 

Using these basis vectors, we can derive selection rules for the magnetic scattering using the following form for the structure factor:
\begin{equation}
\mathscr{F}(\mathbf{Q})=f_S\sum_i \mathbf{M}_i e^{i\mathbf{Q}\cdot r_i}
\label{eq:struct}
\end{equation}
where $\mathbf{Q}$ is the reciprocal lattice vector, $\mathbf{M}_i$ is the ordered moment on the Ir ion at position $r_i$ and the prefactor $f_S=e^{i\pi(h+k)}+e^{i\pi(h+l)}+e^{i\pi(k+l)}$ is the structure factor for the face-centered orthorhombic lattice.
For the (h,0,l) plane, the $F$ and $A$ basis vectors contribute to the $h+l=4n$ peaks with $A$ having very weak peaks at $12n$. The $G$ and $C$ structures, on the other hand, contribute to $h+l=4n+2$ with vanishingly small peaks at 6n for $G$ basis vector.  Figure \ref{fig:spiral}A shows the normalized intensity along the $(0,0,l)$ plane at $\unit[4]{T}$. The absence of a diffraction peak at $(0,0,6)$ and $(0,0,18)$ agrees with what is expected of a $G$ type magnetic order. However, this magnetic order along would not explain all of our data since we also see a change in the structural position $(0,0,4n)$ with an applied field. We will present some symmetry and energy arguments in the next section, that let us to conclude that our data is best explained by a combination of $G$ and $F$ basis vectors.
	
\begin{table}
\newcolumntype{x}[1]{%
>{\centering\hspace{0pt}}p{#1}}%
{\setlength{\extrarowheight}{5pt}
\begin{tabular}{x{1cm} x{1cm} x{1cm} x{1cm}}
\hline \hline

Site & \multicolumn{3}{c}{Coordinates} \tabularnewline
\hline
1 & \multicolumn{3}{c}{(\nicefrac{1}{8}, \nicefrac{1}{8}, $z$)} \tabularnewline
2 & \multicolumn{3}{c}{(\nicefrac{1}{8}, \nicefrac{5}{8}, $\nicefrac{3}{4}-z$)} \tabularnewline
3 & \multicolumn{3}{c}{(\nicefrac{3}{8}, \nicefrac{3}{8}, $1-z$)} \tabularnewline
4 & \multicolumn{3}{c}{(\nicefrac{3}{8}, \nicefrac{7}{8}, $\nicefrac{1}{4}-z$)} \tabularnewline

\hline
\multicolumn{4}{c}{Basis Vectors} \tabularnewline
\hline
\multicolumn{4}{c}{F = $\F$ G = $\G$ A = $\A$  C = $\C$ } \tabularnewline
\hline \hline 
\end{tabular}}
\caption{Ir positions in the orthorombic unit cell and basis vectors for a magnetic structure with propagation vector ${\bf q}= (0,0,0)$.}
\label{position}
\end{table}


\begin{figure}[ht]
    	{\includegraphics[width=7cm]{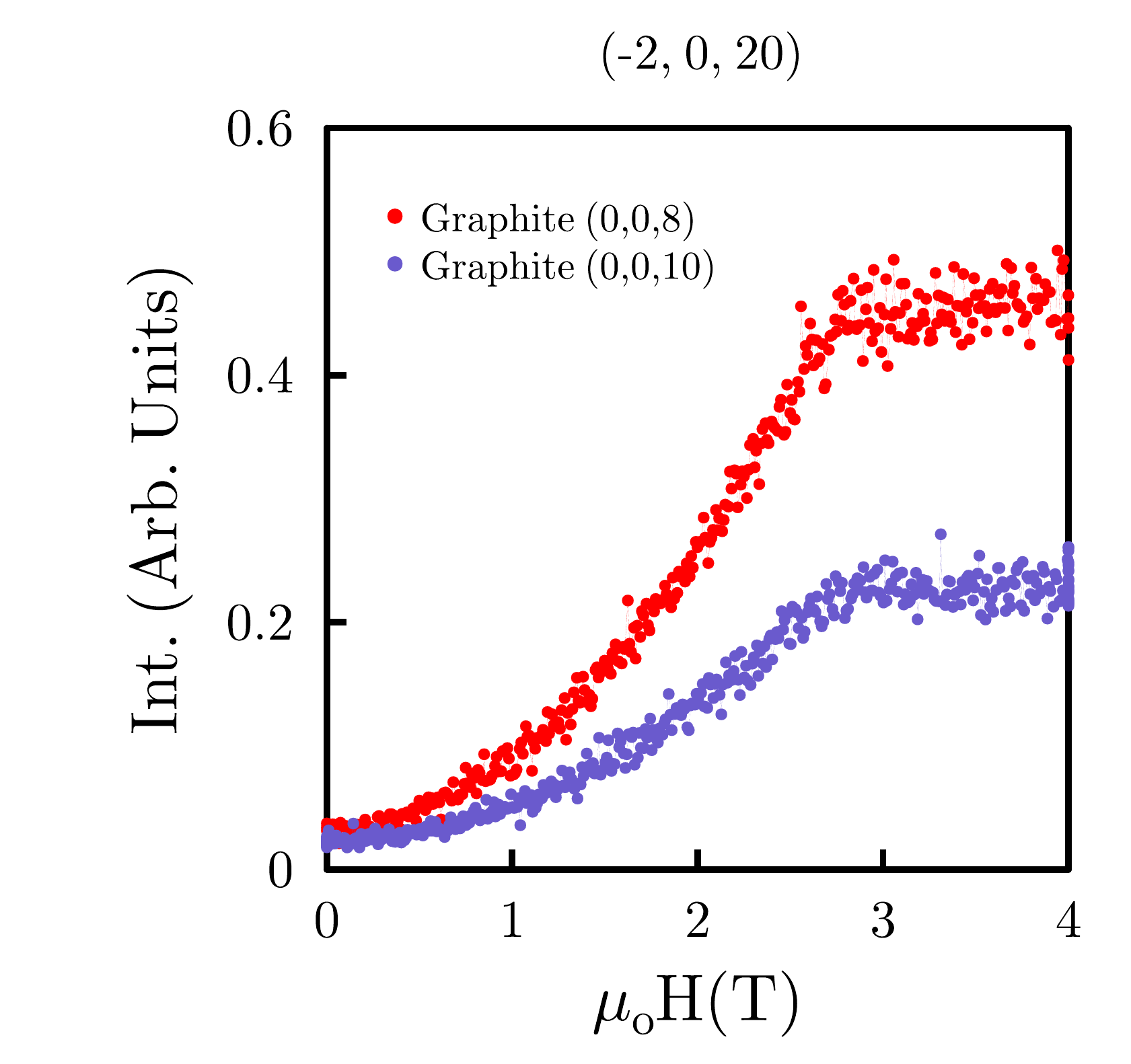}
  	\caption{ {\bf Polarization study of a $h+l=12n\pm2$ peak.} The polarization measurements were performed using the $(0,0,8)$ and $(0,0,10)$ directions of a graphite analizer. }
  	\label{fig:pol}}
	\end{figure}

To understand why $\Psi_V$ grows with rising field or decreasing temperature, we note that the cross section for resonant x-ray magnetic scattering at a given $q$-vector is proportional to $\sum_i e^{\mathrm{i} q\cdot \vec{r}_i}(\hat{\epsilon}_{out}\times \hat{\epsilon}_{in})\cdot \hat{m}_i \sigma_i$, where $\hat{\epsilon}_{out (in)}$ denotes the polarization state of the scattered (incident) beam, $\hat{m}_i$ is a unit vector along the magnetic moment at site $i$, and the sum in $i$ runs over the magnetic unit cell. $\sigma_i$ is a quantity proportional to the local imbalance of magnetic up-down states, and thus proportional to the magnetic moment at site $i$. 

\begin{table}[!hb]
\begin{center}
{\renewcommand{\arraystretch}{1.5}
\begin{tabularx}{0.45\textwidth}{ >{\setlength\hsize{1\hsize}\centering}c|>{\setlength\hsize{1\hsize}\centering}X >{\setlength\hsize{1\hsize}\centering}X }
\hline \hline\
{\bf $\hat{\epsilon}_{in}$ $\times$ $\hat{\epsilon}_{out}$} & {\bf $\hat{m}=\hat{a}$} & {\bf $\hat{m}=\hat{b}$} \tabularnewline 
\hline
{\bf $\hat{\pi}$ $\times$ $\hat{\sigma}= \hat{k}_{in}$} & $\hat{k}_{in} \cdot \hat{a} \sim cos(\theta)$ & {\bf  $\hat{k}_{in} \cdot \hat{b} = 0$} \tabularnewline 
{\bf $\hat{\pi}$ $\times$ $\hat{\pi}= \hat{\sigma}$} & $\hat{\sigma} \cdot \hat{a} = 0$ & {\bf  $\hat{\sigma} \cdot \hat{b} \sim 1 $} \tabularnewline 

\hline \hline 
\end{tabularx}}
\caption{The incident beam is $\pi$-polarized. The magnetic field is alogn the $\hat{b}$-axis, in this geometry, parallel to the $\sigma$ direction. }
\label{polarization}
\end{center}
\end{table}

For a $\pi$-polarized incoming beam, magnetic resonant scattering occurs in both the $\pi$-$\pi$ and $\pi$-$\sigma$ channels. The product  $\hat{\epsilon}_{\pi}\times\hat{\epsilon}_{\sigma}=\hat{k}_{in}$ and $\hat{\epsilon}_{\pi}\times\hat{\epsilon}_{\pi}=\hat{\epsilon}_\sigma$ so that the only contribution of the moment to the scattering intensities comes from the parallel projection along the incoming beam, $\hat{k}_{in}$, and/or along the normal polarization direction of the beam, $\hat{\epsilon}_\sigma$.  Typically, azimuthal scans are performed to infer the moment's direction by keeping the set-up in the scattering condition and rotating the sample around the scattering wavevector, $\hat{Q}=\hat{k}_{out}-\hat{k}_{in}$. The projection of the moment onto the fixed directions,  $\hat{\epsilon}_\sigma$ and/or $\hat{k}_{in}$, varies depending on the azimuthal angle $\Psi$, with maximum scattering for small angle between $\hat{m}$ and the fixed directions, and minimum scattering when they are perpendicular. Unfortunately, we were not able to perform azimuthal scans to determine the moment's direction and the relative phase between the possible basis vector. However, by probing the polarization of the outcoming beam and by studying the $q$-dependence along the $(0,0,l)$ direction, we can infer information about the moment's orientation. 

Figure \ref{fig:spiral}A shows the $q$-dependence along the $(0,0,l)$ direction at \unit[4]{T}. This data can be directly compared with the expected intensity behavior of the $\pi$-$\pi$ and $\pi$-$\sigma$ channels with moments along the $\hat{a}$ and $\hat{b}-$axis listed on Table \ref{polarization}. Since the scattered intensity decreases as the $q$-vector is increased, we conclude that the moments are mostly along $\hat{a}$ and that the major contribution to the intensity is in the $\pi$-$\sigma$ channel.
 
Polarization analysis were performed using the $(0,0,8)$ and $(0,0,10)$ directions of a graphite crystal with structural factor $F_{(0,0,8)}\sim4.40$ and $F_{(0,0,10)}\sim3.24$ and intensity ratio: 
\begin{equation}
\frac{|F_{(0,0,8)}|^2}{|F_{(0,0,10)}|^2} \sim 2
\end{equation}
which is very similar to that observed on the data (Figure \ref{fig:pol}). Since $I=I_{\pi\sigma}+cos\phi \cdot I_{\pi\pi}$, where $\phi$ is the polarization angle, and we observed no $\phi$ dependence in the polarization data (except for what's expected from the intensity ratio), we conclude that most of the intensity comes from the $\pi-\sigma$ channel which agrees with the $q$-dependence and the energy arguments presented below.

\subsection{Linear coupling of a zigzag spin pattern to a uniform magnetic field in $\beta$-Li$_2$IrO$_3$ }
\subsubsection{Symmetry analysis}
To look for possible coupling between a spatial spin pattern and an external field, such as a spatial modulation of g-factor anisotropy, we first perform a full symmetry analysis of the lattice symmetries. 

Table \ref{sym} show the symmetry transformations of various magnetic configurations under the symmetries of the hyperhoneycomb lattice of $\beta$-Li$_2$IrO$_3$, together with time reversal. We find that an external field couples only to a particular configuration, denoted as Zigzag order. The zigzag order of $S^a$ ($S^b$) spins couples linearly to a uniform magnetic field along the $\hat{b}$ ($\hat{a}$) axis.

\subsubsection{Microscopic mechanisms: spatially-varying local susceptibility}
Consider the environment of an iridium $S=1/2$ site. The local environment of the oxygen octrahedra sets the $g$-factor anisotropy. The symmetry analysis above shows that it is possible for the material to have a spatially modulated $g$-factor with off-diagonal terms, akin to that of Sr$_2$IrO$_4$.  However, here the iridium-oxygen octahedra do not exhibit significant rotations. We therefore expect the magnitude of the site-modulated off-diagonal $g^{ab}$ term to here be quite small, less than a few percent.

Once magnetic correlations sample the environment beyond a single Ir site, however, the lattice symmetries immediately come into play. In particular, the local orientation of a zigzag chain produces a preferred local coordinate system for the renormalized magnetic susceptibility. This local coordinate system alternates among sites, in precisely the zigzag pattern. 

One can model this effect as a spatial modulation in a local magnetic susceptibility tensor. The tensor is diagonal in the $\hat{a},\hat{b},\hat{c}$ axis, but has an additional off-diagonal component, 
\begin{equation}
 \chi^{ab}=-\chi^{ba} = (-1)^{\text{zigzag chain}}
\end{equation}
The sign of this component alternates upon crossing a $\hat{c}$-axis bond, ie it alternates between successive zigzag chains.

The effect of this coupling is to produce a zigzag-a (zigzag-b) configuration, together with net $\hat{b}$ ($\hat{a}$) alignment, when a magnetic field is applied along the $\hat{b}$ ($\hat{a}$) axis. 

\begin{table}
\begin{tabular}{|l|c|c|c|c|c|c|}
\hline
\multicolumn{7}{ |c| }{Symmetry transformations} \\
\hline
  &  & $R_a$  & $R_b$  & $R_c$  & $I$  & $T$ \\
 \hline 
\multirow{3}{*}{Uniform field (FM)} 
 & $S^a$  & $+$  & $-$  & $-$  & $+$  & $-$  \\ \cline{2-7}
 & $S^b$  & $-$  & $+$  & $-$  & $+$  & $-$  \\ \cline{2-7}
 & $S^c$  & $-$  & $-$  & $+$  & $+$  & $-$  \\ \hline
\multirow{3}{*}{Stripy  } 
 & $S^a$  & $+$  & $-$  & $-$  & $-$  & $-$  \\ \cline{2-7}
 & $S^b$  & $-$  & $+$  & $-$  & $-$  & $-$  \\ \cline{2-7}
 & $S^c$  & $-$  & $-$  & $+$  & $-$  & $-$  \\ \hline
 \multirow{3}{*}{Zigzag } 
 & $S^a$  & $-$  & $+$  & $-$  & $+$  & $-$  \\ \cline{2-7}
 & $S^b$  & $+$  & $-$  & $-$  & $+$  & $-$  \\ \cline{2-7}
 & $S^c$  & $+$  & $+$  & $+$  & $+$  & $-$  \\ \hline
 \multirow{3}{*}{Neel} 
 & $S^a$  & $-$  & $+$  & $-$  & $-$  & $-$  \\ \cline{2-7}
 & $S^b$  & $+$  & $-$  & $-$  & $-$  & $-$  \\ \cline{2-7}
 & $S^c$  & $+$  & $+$  & $+$  & $-$  & $-$  \\ \hline
  \multirow{1}{*}{Noncoplanar spiral} 
 &    & $-$  & $0$  & $0$  & $0$  & $-$  \\ \hline
\end{tabular}
 \label{sym}
\caption{Transformation rules for $g$-factor anisotropies and various magnetic orders under all $\beta$-Li$_2$IrO$_3$ lattice symmetries. The symmetry generators are: $\pi$ rotations $R$ around the orthorhombic axes $a,b,c$, centered at a $c$-bond midpoint; inversion centers $I$ at the midpoint of $d$-bonds; and time-reversal $T$. The space group Fddd also contains glide reflections, which are generated by $R\times I$. The symbols $+,-,0$ denote that a configuration with a given spin orientation $S^{a,b,c}$ is respectively even, odd, or fully-breaking under the symmetry. The zigzag order of $S^a$ ($S^b$) spins couples linearly to a uniform magnetic field along the $\hat{b}$ ($\hat{a}$) axis. }
\end{table}

\subsection{Possible vestigial orders of the $\beta$-Li$_2$IrO$_3$ spiral}
\subsubsection{Symmetries of the spiral order}
Let us consider the spiral order observed in  $\beta$-Li$_2$IrO$_3$. We assume it is incommensurate. The observed spiral has wavevector along $a$. Its basis vectors all belong to a single irreducible representation $\Gamma_4$ at this wavevector, consisting of basis vectors $A_x,C_y,F_z$, where $x,y,z$ here refer to spin directions along the orthorhombic axes $a,b,c$ respectively. Here $F$ is uniform, and is $\pi/2$ out of phase with the nonuniform basis vectors $A$ and $C$ which are present. For the sites given in the order above, $C=(++--)$ and $A=(+-+-)$. 

Now consider its lattice symmetries. As seen in the table, the operations $I, G_a, G_b, G_c, R_b, R_c$ are completely broken, in that they each take the spin configuration into a completely different configuration, which remains different even up to an overall spin flip. Thus, the product of each of these symmetry operations with time reversal $T$ also remains broken.  However, the remaining operation $R_a$ takes each magnetic moment precisely to its opposite. So its product with time reversal, $T R_a$, is preserved as a symmetry of the spiral. 

This symmetry analysis assumes that translations are fully broken along the spiral wavevector.  A commensurate spiral can have a few additional symmetry operations, associated with its very large commensurate unit cell. However, these symmetry operations require fine-tuning of the overall phase of a commensurate order. Lacking any experimental evidence for such phase-locked commensurate ordering, we here focus on the generic case, where the wavevector is incommensurate or, if commensurate, with generic overall phase.

\subsubsection{Symmetry implications of the spiral order}
From the analysis above, we see that the spiral ordering preserves only a single space-group symmetry operation: $T R_a$, the product of time reversal and a rotation by $\pi$ around the crystallographic $a$-axis passing through a $c$-bond midpoint. Aside from the lattice translations along $b$ and $c$, the spiral order breaks all crystal symmetries except for the single symmetry $T R_a$. 

It is therefore possible for the system to simutaneously develop an order parameter for any order which preserves these symmetries, namely $T R_a$ as well as $b,c$ translations. Together with the $a$-axis spiral, the following $q=0$ orders are therefore symmetry allowed. (1) Ferromagnetic alignment along $S^b$ or $S^c$, denoted FM-b,c. (2) C-Stripy order (spins aligned across $z$-type i.e. $c$-type bonds and antialigned elsewhere), with spins again along $S^b$ or $S^c$, denoted Stripy-b,c. (3) Neel order, with spins along $S^a$, denoted Neel-a. (4) C-Zigzag order (spins anti-aligned across $z$-type i.e. $c$-type bonds and aligned elsewhere), with spins again along $S^a$ denoted C-Zigzag-a.  For convenience, we only consider the ``C'' versions of the stripy and zigzag orders, and drop the ``C'' prefix henceforth.

\subsubsection{Bragg peak signatures}
These orders can be distinguished by their Bragg peak signatures. The Bragg peak selection rules are independent of the spin orientation. For clear comparison with experiment, let us focus on the peaks at $k=0$, i.e.\ the  (h,0,l) plane. In this plane, the (h,0,l) Bragg peaks can be parametrized by two arbitrary integers $m,n$, as follows:

   (1) FM or structural peaks: $(2m, 0, 4n+2m)$, with strong peaks at  $(2m, 0, 12n+6m)$.
   
   (2) Stripy peaks: $(2m, 0, 4n+2m+2)$, with strong peaks at  $(2m, 0, 12n+6m+6)$.  
   
   (3) Neel peaks: $(2m,0,12n +6m \pm 4)$. 
   
   (4) Zigzag peaks: $(2m,0,12n +6m \pm 2)$.

\subsubsection{Energetics of possible vestigial orders}
Based on the known information on the Hamiltonian of the zero-field spiral order, which has dominant FM Kitaev exchange, we can estimate the relative energies of these competing vestigial orders. The Neel state is disfavored due to the strong FM exchange along all nearest neighbor bonds. For the stripy pattern, where spins are aligned only along $c$-bonds, the FM $b$-axis Kitaev coupling on these $c$-bonds would favor Stripy-b. For the zigzag pattern, where spins are aligned only along $x,y$-bonds, the FM $a,c$-axis Kitaev coupling on these bonds would favor Zigzag-a. 

 Finally by observing the known magnetic susceptibility, which is much stronger along $b$ than along $a$, we note that for the FM patterns, the FM-b is observed to be more easily stabilized than FM-c. The candidate phases with likely lower energy are thus as follows: (1) FM-b; (2) Stripy-b; and (3) Zigzag-a.
 
 The Zigzag-a configuration is linearly coupled to FM-b. Applying an external $b$-field would then be expected to disfavor Stripy-b as well as the more energetically-costly vestigial possibilities, while favoring FM-b together with Zigzag-a.

\subsubsection{Quantitative estimate of the Kitaev-Heisenberg exchange}
\label{sec:KHest}
A quantitative estimate of the energetics of the field-induced vestigial order can shed light on the  Kitaev-based model for \blio\ at zero field. Consider the vestigial order parameter $\Psi_{\textrm{V}}$, defined as the magnitude (in units of $\hbar/2$) of the local spin whose magnetic moment orientation is locked to the lattice $ b \pm a$ directions. 
Its energy per site  $E$ is given by an expectation value of the \blio\ Hamiltonian supplemented by a Zeeman term for the applied field $H$. Using the $K$-$J$-$I_c$ model Hamiltonian for \blio\ given in Ref.\cite{kimchi_unified_2015}, one finds that the pseudo-dipolar $I_c$ parameter drops out, resulting in the expression:
\begin{align*}
E =& - \frac{g \mu_B}{2} (\vec{H}\cdot \hat{b}) {\rm cos}\theta \Psi_{\textrm{V}} \\
     &+\frac{J}{8} (3 {\rm cos}^2\theta + {\rm sin}^2\theta)(\Psi_{\textrm{V}})^2\\
     &+ \frac{K}{8} ({\rm cos}^2\theta +{\rm sin}^2\theta ) (\Psi_{\textrm{V}})^2\\
   E_v =&- 0.047 H(T) \Psi_{\textrm{V}}+ (0.292 J + 0.125 K) (\Psi_{\textrm{V}})^2
\end{align*}

With $g=2$ and $\theta = \tan^{-1} 1/\sqrt{2} \approx 0.2 \pi$ corresponding to the  $ b \pm a$ lattice locking of $\Psi_{\textrm{V}}$. Constraints on the magnitude of the ferromagnetic Kitaev ($K<0$) and antiferromagnetic Heisenberg ($J>0$) interactions can be derived by comparing $E_V$ to the energy per site of the incommensurate spiral order, estimated via mean-field from $T_I =\unit[38]{K}$ to be about $E_I \approx \unit[-1.6]{meV}$. 
Our measurement, showing that a small $H=\unit[2.8]{T}$ field is sufficient for destroying the incommensurate order in favor of a saturated vestigial order $\Psi_{\textrm{V}}$, implies a high degree of fine tuning between $E_V$ and $E_I$: taking $\Psi_{\textrm{V}}=1$ at $H=\unit[2.8]{T}$, we find that the Kitaev and Heisenberg interactions must obey the following constraint,
\begin{equation}
K \approx\unit[-11.8]{meV} -\unit[2.3]{J}
\end{equation}
The few-Tesla instability of the incommensurate spiral in favor of the vestigial order thus directly implies that the Kitaev interaction must be significantly larger than the Heisenberg exchange, and indeed must dominate the physical response of the material at both zero and finite applied fields.

\subsubsection{Hyperhoneycomb lattice symmetries}
The space group of the hyperhoneycomb lattice is Fddd, number 70.
The lattice symmetries consist of the following operations. Together with the primitive translation vectors of the face-centered Bravais lattice, there are two-fold rotations through c-bonds, and inversion centers and glide planes through d-bonds, as follows:

(1) There are two-fold rotations $R$, of three types, corresponding to each of the three orthorhombic axes $a,b,c$. Each is a rotation by $\pi$ around one of the orthorhombic axes. The rotation axis passes through a midpoint of a c-bond.

(2) There are inversion centers $I$, at the midpoint of d-bonds. (The d-bonds are the remaining ``diagonal'' bonds, which form the zigzag chains; they carry Kitaev labels x or y.)  

(3) There are glide operations $G$, again of three types corresponding to the three orthorhombic axes $a,b,c$. The $c$-type glide plane involves reflection across the $(a,b)$ plane normal to $c$, together with translation by half the primitive lattice vector normal to $c$, i.e.\ translation by $(1/4,1/4,0)$ in orthorhombic coordinates. The $a$ and $b$ glide planes are similarly defined. For all of the glide operations, the reflection plane passes through a midpoint of a d-bond. 

The midpoint of a c-bond is related to the midpoint of a d-bond through a translation by vectors such as $(1/8,1/8,1/8)$, up to component-wise $\pm$ signs. Note that the standard setting of Fddd (origin choice 2) places the origin at the midpoint of a d-bond. 

The primitive unit cell of the lattice contains 4 sites. Taking the origin to be at a c-bond midpoint, we can write the coordinates of the first two sites, across this c-bond, as $(0,0,\pm1/12)$. The other two sites have coordinates $(1/4,1/4,1/4\pm1/12)$. We can shift the origin to a d-bond midpoint by shifting the sites by $-(1/8,1/8,1/8)$. 

Finally, we note the relations among the symmetry operations. Each glide operation is related to the respective ($a,b,c$) rotation, through a product with the inversion symmetry. Moreover, each pair of rotations generates the third. Thus the full lattice symmetries can be generated from the translations, an inversion center and two rotations.

%

\end{document}